\documentstyle[axodraw]{article}

\def\ltap{~\raisebox{-.4ex}{\rlap{$\sim$}} \raisebox{.4ex}{$<$}~}

\newcommand{\Rsl}{{\not\!{R}}}
\renewcommand{\thefootnote}{\fnsymbol{footnote}}

\title{\begin{flushright}
\small 
SINP/TNP/98-22\\
{\tt hep-ph/9809493} 
\end{flushright}
{\bf Upper bounds on all $R$-parity-violating
$\lambda\lambda''$ combinations from proton stability}}

\author{
{\sf Gautam Bhattacharyya}%
\thanks{E-mail address: gb@tnp.saha.ernet.in}~~and 
{\sf Palash B. Pal}%
\thanks{E-mail address: pbpal@tnp.saha.ernet.in} \\[2.5mm]
Saha Institute of Nuclear Physics, 1/AF Bidhan Nagar, 
Calcutta 700064, India
}

\date{}

\begin{document}

\maketitle

\begin{abstract}
  
{\small In an $R$-parity-violating supersymmetric theory, we derive
upper bounds on all the $\lambda''_{ijk}\lambda_{i'j'k'}$-type
combinations from the consideration of proton stability, where
$\lambda''_{ijk}$ are baryon-number-violating couplings involving
three baryonic fields and $\lambda_{i'j'k'}$ are
lepton-number-violating couplings involving three leptonic fields.}

\end{abstract}

\vskip 20pt  

\setcounter{footnote}{0}
\renewcommand{\thefootnote}{\arabic{footnote}}

\paragraph*{1. }
In the minimally supersymmetrized standard model (MSSM), the
superpotential contains the following $R$-parity conserving terms: 
        \begin{eqnarray}
\label{W0}
W_0 & = & f_e^{ij} L_i H_d E_j^c + f_d^{ij} Q_i H_d D_j^c
+ f_u^{ij} Q_i H_u U_j^c + \mu H_d H_u \,.
        \end{eqnarray}
Here, $L_i$ and $Q_i$ are SU(2)-doublet lepton and quark superfields;
$E^c_i, U^c_i, D^c_i$ are SU(2)-singlet charged lepton, up- and
down-quark superfields; $H_d$ and $H_u$ are Higgs superfields
responsible for the down- and up-type masses respectively. The
generation indices are assumed to be summed over.

$R$-parity is a discrete symmetry which is defined as $R =
(-1)^{(3B+L+2S)}$, where $B$ is the baryon number, $L$ is the lepton
number and $S$ is the spin of the particle. $R$ is $+1$ for all
standard model particles and $-1$ for their superpartners.  If one
allows $R$-parity violation \cite{rpar}, the most general
superpotential includes the following $L$- and $B$-violating terms:
        \begin{eqnarray}
\label{W'}              
W' & = & {1\over{2}}\lambda_{ijk} L_i L_j E^c_k + 
\lambda'_{ijk} L_i Q_j D^c_k + 
{1\over{2}}\lambda''_{ijk} U^c_i D^c_j D^c_k + 
\mu_i L_i H_u \,.
        \end{eqnarray}
Here $\lambda''_{ijk}$ are $B$-violating while $\lambda_{ijk}$,
$\lambda'_{ijk}$ and $\mu_i$ are all $L$-violating couplings.
Considering the antisymmetry in the first (last) two flavor indices in
$\lambda$ ($\lambda''$), namely
        \begin{eqnarray}
\lambda_{ijk}=-\lambda_{jik} \,, \qquad 
\lambda''_{ijk}=-\lambda''_{ikj}\,, 
\label{antisymm}
        \end{eqnarray}
there are 48 additional parameters. These are constrained from
various experimental searches~\cite{review}.

\paragraph*{2. }
Simultaneous presence of $B$- and $L$-violating couplings drive proton
decay.  Therefore, in an $R$-parity-violating ($\Rsl$) theory, what can
be derived from proton decay are, for example, bounds on $\lambda''$
correlated with any of the $L$-violating couplings. As mentioned
before, $L$-violating sources are of 3 types.  The couplings
$\lambda'_{ijk}$ constitute one type of source, and the correlated
bounds in this case exist in the literature \cite{HK,SmVi96}. At the
tree level the bounds apply only to a select set of the couplings, and
one obtains \cite{HK}, assuming superpartner masses around 1 TeV,
        \begin{equation} 
\lambda'_{11k} \lambda''_{11k} \ltap 10^{-24},
\label{lpldp}
                \end{equation}
where $k = 2, 3$. At the one-loop level, one can always find at least
one diagram in which any $\lambda'_{ijk}$ in conjunction with any
$\lambda''_{lmn}$ contributes to proton decay \cite{SmVi96}. It
follows that for superparticle masses of order 1~TeV,
        \begin{equation}
\lambda'_{ijk} \lambda''_{lmn} \ltap 10^{-9}.  
        \end{equation}
If one admits tree level flavor-changing squark mixing, the bounds
are strengthened by two orders of magnitude. 

Recently, contributions to proton decay originating from the
$L$-violating parameters $\mu_i$ in conjunction with the $\lambda''$'s
have been investigated \cite{BP1}. Here, the diagram at the tree level
produces a constraint, for an exchanged scalar mass of 1 TeV,
\begin{eqnarray} \lambda''_{112} \epsilon_l < 10^{-21} \,,
\end{eqnarray} where $\epsilon_l=\mu_l/\mu$, with $\mu$ assumed to be
of order 1 TeV. Constraints on the other
$\lambda''_{ijk}\epsilon_l$-type combinations originate from loop
diagrams and hence are weaker. They are typically of order $10^{-10}$
to $10^{-14}$, always assuming the superparticle masses of order 1
TeV.

\paragraph*{3. }
{\em The aim of the present paper} is to examine the other source of
lepton number violation, viz.\ $\lambda_{ijk}$, and derive bounds on
$\lambda''_{ijk}\lambda_{i'j'k'}$ products with any choices of flavor
indices.

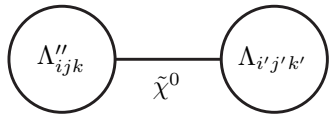
\begin{figure}
\begin{center}
\begin{picture}(140,50)(40,30)
\SetWidth{1.1}
\BCirc(70,50){20}
\Text(70,50)[]{$\Lambda''_{ijk}$}
\Line(90,50)(130,50)
\Text(110,45)[t]{$\tilde{\chi}^0$} 
\BCirc(150,50){20}
\Text(150,50)[]{$\Lambda_{i'j'k'}$}
\end{picture}
\end{center}
\caption[]{\small\sf Generic structure of diagrams involving
$\lambda''$ and $\lambda$ couplings that lead to proton
decay.}\label{f:generic}
\end{figure}

The fact that $\lambda$ and $\lambda''$ together can drive proton
decay has been noted before \cite{LoPa97} in the context of an
extended gauge model. The idea however applies to any general
framework of $R$-parity violation, for example, as the one in the
present paper.  We will note as we proceed further that any one of the
nine $\lambda''$-couplings in association with any one of the nine
$\lambda$-couplings can contribute to proton decay at one- or two-loop
order if not at the tree level. Fig.~\ref{f:generic} shows a generic
diagram involving two blobs. The $\Lambda''_{ijk}$-blob on one side
represents either a tree or a one-loop diagram containing a
$\lambda''_{ijk}$-vertex with external quark lines. The
$\Lambda_{i'j'k'}$-blob on the other side contains a
$\lambda_{i'j'k'}$-vertex with external lepton lines through a tree or
a one-loop diagram. The two blobs are connected by virtual
neutralinos. The amplitude of this generic diagram can be written as
        \begin{eqnarray}
G_{\Rsl} \simeq 
{\Lambda''_{ijk} \Lambda_{i'j'k'} \over {m_{\tilde{\chi}^0}}} \,,
\label{GBviol}
        \end{eqnarray}
where $m_{\tilde{\chi}^0}$ is the mass of the exchanged
neutralino. The maximum contribution is expected to come from the
exchange of the lightest neutralino, which we assume to be
predominantly a gaugino (the Higgsino exchanged graphs will be
suppressed by light masses).  The next task is to decipher the
explicit structures of the two blobs $\Lambda''$ and $\Lambda$, that
involve all possible combinations of flavor indices associated with
$\lambda''$ and $\lambda$, in a case by case basis.

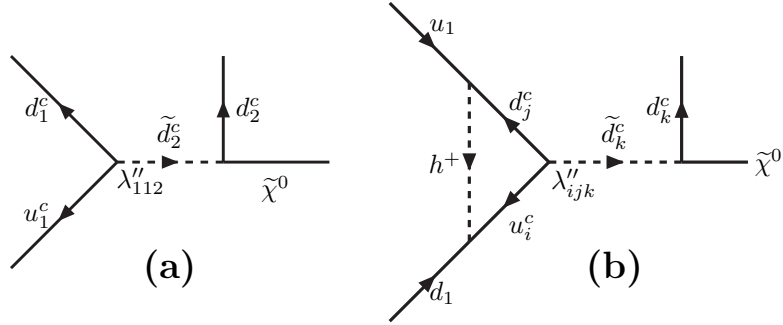
\begin{figure}
\begin{center}
\begin{picture}(120,100)(10,0)
\SetWidth{1.1}
\ArrowLine(50,50)(10,10)
\Text(25,30)[r]{$u^c_1$}
\ArrowLine(50,50)(10,90)
\Text(25,70)[r]{$d^c_1$}
\Text(50,48)[tl]{$\lambda''_{112}$}
\DashArrowLine(50,50)(90,50){3}
\Text(70,58)[b]{$\widetilde d^c_2$}
\ArrowLine(90,50)(90,90)
\Text(95,70)[l]{$d^c_2$}
\Line(90,50)(130,50)
\Text(110,45)[t]{$\widetilde \chi^0$}
\Text(70,10)[]{\Large\bf (a)}
\end{picture}
\qquad
\begin{picture}(135,100)(-10,0)
\SetWidth{1.2}
\ArrowLine(50,50)(20,20)
\Text(35,30)[tl]{$u^c_i$}
\ArrowLine(-10,-10)(20,20)
\Text(5,5)[tl]{$d_1$}
\ArrowLine(50,50)(20,80)
\Text(35,70)[bl]{$d^c_j$}
\ArrowLine(-10,110)(20,80)
\Text(5,100)[bl]{$u_1$}
\DashArrowLine(20,80)(20,20){3}
\Text(17,50)[r]{$h^+$}
\DashArrowLine(50,50)(100,50){3}
\Text(75,58)[b]{$\widetilde d^c_k$}
\Text(51,40)[bl]{$\lambda''_{ijk}$}
\ArrowLine(100,50)(100,90)
\Text(97,70)[r]{$d^c_k$}
\Line(100,50)(125,50)
\Text(128,50)[l]{$\widetilde \chi^0$}
\Text(75,10)[]{\Large\bf (b)}
\end{picture}
%
%

\end{center}
\caption[]{\small\sf The $B$-violating blob $\Lambda''_{ijk}$ for
various combinations of the indices.  $\tilde\chi^0$ is the neutralino
coupled to the blob.}\label{f:Lambda''}
\end{figure}
\paragraph{4. }
We start with the evaluation of $\Lambda''_{ijk}$ for various
combinations of the indices. Note that, due to the antisymmetry of the
indices mentioned in Eq.\ (\ref{antisymm}), we can always take
$k\neq3$. Among the independent couplings now, we can distinguish two
cases. This part is very similar to the discussion appearing in our 
earlier work~\cite{BP1}.

\subparagraph{Case a)} 
For $\Lambda''_{112}$, we obtain a tree graph, which is shown in
Fig.~\ref{f:Lambda''}a. The strength of the blob is given by
	\begin{eqnarray}
\Lambda_{112}^{\prime\prime\,(a)} \approx {g\lambda''_{112} \over
m^2_{\tilde s^c}} \,. 
	\end{eqnarray}
%

\subparagraph{Case b)}
The other $\lambda''_{ijk}$'s cannot appear in tree diagrams because
they would involve at least one heavy quark ($c$, $b$ or $t$) in the
outer legs. However, they couple through loop diagrams involving the
exchange of charged scalars. Due to the $\mu$-term, the physical
charged Higgs $h^+$ is a combination of $H_u^+$ and $H_d^+$.  The
diagram involving $h^+$ is shown in Fig.~\ref{f:Lambda''}b where
flavor violation occurs via Cabibbo-Kobayashi-Maskawa (CKM)
projections.  This diagram is infinite. In the supersymmetric limit,
it cancels with a similar diagram containing the longitudinal
$W$-boson in place of the charged Higgs. Since supersymmetry is
broken, the sum of the two diagrams gives a finite quantity which
depends logarithmically on the masses of these two scalars. For a
charged Higgs mass of order 1\,TeV that we assume in this paper, we
can safely omit this logarithm for an order-of-magnitude estimate,
yielding
\begin{eqnarray} 
\Lambda''_{ijk} \approx {f_u^i f_d^j \over
16\pi^2} \; (V^*_{i1} V_{1j}) \; {g\lambda''_{ijk} \over m^2_{\tilde
d^c_k}} \,.  
\end{eqnarray}
%

\paragraph{5. }
We now discuss the lepton-number violating blob
$\Lambda_{i'j'k'}$. Here, there are three cases which should be
distinguished.

\begin{figure}[b]
\begin{center}
\begin{picture}(120,100)(0,0)
\SetWidth{1.1}
\Line(0,50)(30,50)
\Text(15,45)[t]{$\widetilde \chi^0$}
\ArrowLine(30,90)(30,50)
\Text(25,70)[r]{$\ell^c_{k'}$}
\DashArrowLine(30,50)(70,50){3}
\Text(50,58)[b]{$\widetilde\ell^c_{k'}$}
\ArrowLine(100,80)(70,50)
\Text(95,70)[l]{$\ell_{j'}$}
\ArrowLine(100,20)(70,50)
\Text(95,30)[l]{$\nu_{i'}$}
\Text(75,50)[l]{$\lambda_{i'j'k'}$}
\Text(50,10)[]{\Large\bf (a)}
\end{picture}
\qquad
\begin{picture}(120,100)(100,0)
\DashArrowLine(142.5,30)(130,50){2}
\Text(135,40)[r]{$\tilde\ell_3$}
\DashArrowLine(142.5,30)(155,10){2}
\Text(144,18)[r]{$\tilde\ell^c_3$}
\SetWidth{1.1}
\Line(100,50)(130,50)
\Text(115,45)[t]{$\widetilde \chi^0$}
\ArrowLine(130,50)(155,90)
\Text(140,70)[br]{$\ell_3$}
\Photon(155,90)(180,50){2}{6}
\Text(172,70)[l]{$W$}
\ArrowLine(180,50)(155,10)
\Text(172,30)[l]{$\ell_2$}
\GCirc(142.5,30){3}{0}
\ArrowLine(155,90)(185,90)
\Text(170,95)[b]{$\nu_3$}
\ArrowLine(210,50)(180,50)
\Text(195,55)[b]{$\nu_2$}
\ArrowLine(185,10)(155,10)
\Text(170,15)[b]{$\nu_1$}
\Text(155,8)[t]{$\lambda_{123}$}
\Text(115,10)[]{\Large\bf (b)}
\end{picture}
\qquad
\begin{picture}(110,85)(50,0)
\SetWidth{1.2}
\Line(50,50)(70,50)
\Text(60,45)[t]{$\widetilde \chi^0$}
\ArrowArc(95,50)(25,90,180)
\Text(73,70)[r]{$\ell_{k'}^c$}
\ArrowArcn(95,50)(25,90,0)
\Text(117,70)[l]{$\ell_{k'}$}
\GCirc(95,75){3}{0}
\DashArrowLine(70,50)(120,50){3}
\Text(95,45)[t]{$\widetilde\ell_{k'}^c$}
\ArrowLine(150,50)(120,50)
\Text(138,56)[b]{$\nu_{i'}$}
\Text(120,48)[t]{$\lambda_{i'k'k'}$}
\Text(95,10)[]{\Large\bf (c)}
\end{picture}
\end{center}
\caption[]{\small\sf The $L$-violating blob $\Lambda_{i'j'k'}$ for
various combinations of the indices. $\tilde\chi^0$ is the neutralino
coupled to the blob.}\label{f:Lambda}
\end{figure}
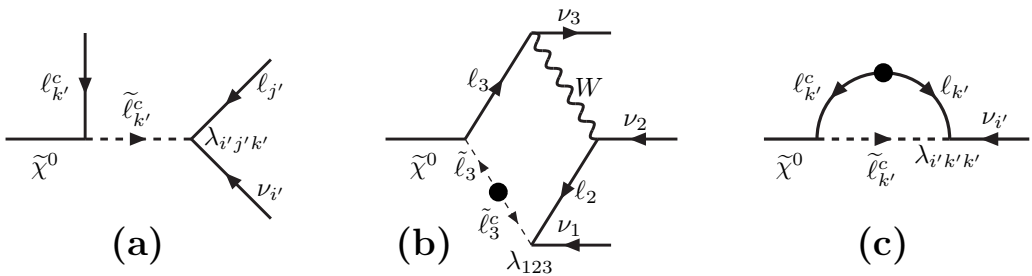
\subparagraph{Case a)}
Six of the nine different $\lambda_{i'j'k'}$'s, characterized by
$k'\neq3$, contribute to the blob at the tree level as shown in
Fig.~\ref{f:Lambda}a. The strength of this blob can be written as 
	\begin{eqnarray}
\Lambda_{i'j'k'}^{(a)} \approx {g\lambda_{i'j'k'} \over
m^2_{\tilde \ell^c_{k'}}} \,, \qquad (k'\neq3).
	\end{eqnarray}
%

\subparagraph{Case b)}
For $\Lambda_{123}$, the dominant contribution comes through the box
shown in Fig.~\ref{f:Lambda}b. Again, an order of magnitude estimate
for the strength of this blob yields
	\begin{eqnarray}
\Lambda_{123}^{(b)} \approx {g^3\lambda_{123} \over
16\pi^2 M_W^2} \, {m_\tau \over m_{\tilde \tau}} \,.
	\end{eqnarray}
Here we have parametrized the left-right slepton mixing blob as $m_\tau
m_{\tilde\tau}$, where $m_{\tilde\tau}$ is some kind of an average of
the masses of $\tilde \tau_L$ and $\tilde\tau_R$.

\subparagraph{Case c)}
There is another type of contribution to the blob $\Lambda_{i'j'k'}$
for $j'=k'$ (or equivalently $i'=k'$ on account of the antisymmetry
shown in Eq.\ (\ref{antisymm})). These are self-energy type diagrams
shown in Fig.~\ref{f:Lambda}c. These include some of the couplings
described in case (a), viz., those with $j'=k'=1$ and $j'=k'=2$. For
the remaining case $j'=k'=3$, we do not have the tree diagram
described in case (a) because that would have implied a $\tau$-lepton
in the final state.  The strength of this blob is given by
	\begin{eqnarray}
\Lambda_{i'k'k'}^{(c)} \approx {g\lambda_{i'k'k'} \over
16\pi^2} \, m_{\ell_{k'}} \,.
	\end{eqnarray}
Since these blobs lead to only one particle carrying lepton number in
the final state, the dimension of $\Lambda_{i'k'k'}^{(c)}$ is not the
same as that of $\Lambda_{i'j'k'}^{(a)}$ or $\Lambda_{123}^{(b)}$.
Since such self-energy diagrams involve lower dimensional operators,
the constraints on $\lambda_{i'11}$ and $\lambda_{i'22}$ derived from
these diagrams, despite suffering loop suppressions, happen to be
stronger than those derived from tree diagrams in case~(a).

\paragraph{6. }
We can now put the various combinations of $\Lambda''_{ijk}$ and
$\Lambda_{i'j'k'}$ into Eq.\ (\ref{GBviol}) to obtain the strength of
the baryon and lepton number violating couplings which are responsible
for proton decay. For lepton number violation through case (c), the
effective coupling $G_{\Rsl}$ has mass dimension $-2$, and proton
lifetime is given by
	\begin{eqnarray}
\tau_p = \left( m_p^5 G_{\Rsl}^2 \right)^{-1} \,.
	\end{eqnarray}
In other cases, the final states will have three particles carrying
lepton number, the effective coupling $G_{\Rsl}$ will have mass
dimension $-5$, and the proton lifetime will be given by
	\begin{eqnarray}
\tau_p = \left( m_p^{11} G_{\Rsl}^2 \right)^{-1} \,.
	\end{eqnarray}
We present the bounds on the combinations
$\lambda''_{ijk}\lambda_{i'j'k'}$ in Table~\ref{t:bounds} for
superparticle masses of order 1~TeV. In deriving the bounds, we have
taken $\tau_p$ to be $10^{32}$\,y \cite{pdg}, except for final states
with charged leptons for which a benchmark value of $10^{31}$\,y has
been assumed. Moreover, for our order of magnitude estimates, we have
neglected all final state particle masses and phase space factors, as
in all earlier estimates \cite{review,HK,SmVi96,BP1}. Consideration of
4-body phase space can relax the bounds in the first two rows of
Table\,\ref{t:bounds} by about two orders of magnitude.

\begin{table}
\begin{center}
\begin{tabular}{c||cc|cc}
\hline 
& \multicolumn{2}{c|}{$\lambda''_{112}$} &
\multicolumn{2}{|c}{all other $\lambda''_{ijk}$}  \\
\cline{2-5}
$\lambda_{i'j'k'}$ & Final states & Bounds & Final states & Bounds \\ 
\hline
$i'\neq j' \neq k'$, $k'\neq 3$ & $K^+e^\pm \mu^\mp \bar\nu$ & $10^{-16}$ 
& $\pi^+(K^+) e^\pm \mu^\mp \bar\nu$ & $10^{-5}$--$10^{-7}$ \\ 
&&&&\\
$i'\neq j' \neq k'$, $k'= 3$ & $K^+ + 3\nu$ & $10^{-14}$
& $\pi^+ (K^+) + 3\nu$ & $10^{-3}$--$10^{-5}$\\
&&&&\\
$j'=k'=1$ (or $i'=k'=1$) & $K^+ \bar \nu$ & $10^{-17}$ 
& $\pi^+ (K^+) \bar \nu$ & $10^{-6}$--$10^{-8}$ \\
&&&&\\
$j'=k'=2$ (or $i'=k'=2$) & $K^+ \bar \nu$ & $10^{-20}$ 
& $\pi^+ (K^+) \bar \nu$ & $10^{-9}$--$10^{-11}$ \\
&&&&\\
$j'=k'=3$ (or $i'=k'=3$) & $K^+ \bar \nu$ & $10^{-21}$ 
& $\pi^+ (K^+) \bar \nu$ & $10^{-10}$--$10^{-12}$ \\
\hline
\end{tabular}
\end{center}
\caption[]{\small\sf Decay modes for the proton and bounds derived on
the couplings for all possible combinations of the baryon violating
and the lepton violating vertices. The bounds are on the products
$\lambda''_{ijk}\lambda_{i'j'k'}$. All superpartner masses are assumed
to be 1 TeV. The ranges in the last column indicate variation due to
different CKM projections.\label{t:bounds}}
\end{table}

\paragraph{7. }
To conclude, we have derived new constraints on all products of the
form $\lambda''_{ijk}\lambda_{i'j'k'}$ from proton stability. These
bounds are complementary to other $L$ and $B$ violating products
\cite{SmVi96,BP1} that contribute to proton decay. In most cases, our
bounds are orders of magnitude stronger than the products of upper
bounds on individual couplings \cite{review,nnbar,bcs,bgh}.

\paragraph*{Acknowledgements~:}
GB acknowledges hospitality at the CERN Theory Division where part of
the work was done.


\begin{thebibliography}{[W]}

\bibitem{rpar} G. Farrar and P. Fayet, Phys. Lett. B 76 (1978) 575; S.
Weinberg, Phys. Rev. D 26 (1982) 287; N. Sakai and T. Yanagida,
Nucl. Phys. B 197 (1982) 533; C. Aulakh and R.~N. Mohapatra, Phys.
Lett. B 119 (1982) 136; L. Hall and M. Suzuki, Nucl. Phys. B 231
(1984) 419; J. Ellis et al., Phys. Lett. B 150 (1985) 142; G. Ross and
J.~W.~F. Valle, Phys. Lett. B 151 (1985) 375; S. Dawson, Nucl. Phys. B
261 (1985) 297; R. Barbieri and A. Masiero, Nucl. Phys. B 267 (1986)
679.
  
\bibitem{review} For reviews, see G. Bhattacharyya, hep-ph/9709395,
Nucl. Phys. Proc. Suppl. 52A (1997) 83; H. Dreiner, hep-ph/9707435.
  
\bibitem{HK} I. Hinchliffe and T. Kaeding, Phys. Rev. D 47 (1993) 279.
  
\bibitem{SmVi96} A.~Y. Smirnov and F. Vissani, Phys. Lett.  B 380
(1996) 317. See also, C. Carlson, P. Roy and M. Sher, Phys. Lett. B
357 (1995) 99.

\bibitem{BP1} G. Bhattacharyya and P.~B. Pal, hep-ph/9806214, to
appear in Phys. Lett. B.

\bibitem{LoPa97} H.~N. Long and P.~B. Pal, Mod. Phys. Lett. A 13
(1998) 2355.
  
\bibitem{pdg} Review of Particle Physics, Eur. Phys. J. C 3 (1998) 1.

\bibitem{nnbar} F. Zwirner, Phys. Lett. B 132 (1983) 103; J.~L. Goity
and M. Sher, Phys. Lett. B 346 (1995) 69, Erratum ibid. B385 (1996)
500.

\bibitem{bcs} G. Bhattacharyya, D. Choudhury and K. Sridhar,
Phys. Lett. B 355 (1995) 193.

\bibitem{bgh} V. Barger, G. Giudice and T. Han, Phys. Rev. D 40 (1989)
2987. 

\end{thebibliography}
\end{document}